\documentclass[aps,prb,twocolumn,groupedaddress,showpacs,floatfix]{revtex4}

\usepackage[latin1]{inputenc}
\usepackage{graphicx}
\usepackage{amsmath}
\usepackage{nicefrac}

\begin{document}


\title{Fast readout of a single Cooper-pair box using its quantum
  capacitance}


\author{F. Persson}
\email[]{fredrik.persson@chalmers.se}
\author{C.M. Wilson}
\author{M. Sandberg}
\author{P. Delsing}
\email[]{per.delsing@chalmers.se}
\affiliation{Department of Microtechnology and Nanoscience (MC2),
Chalmers University of Technology, SE-412~96 G\"{o}teborg, Sweden}


\date{\today}

\begin{abstract}
We have fabricated a single Cooper-pair box (SCB) together with an
on-chip lumped element resonator. By utilizing the quantum capacitance
of the SCB, its state can be read out by detecting the phase of a
radio-frequency (rf) signal reflected off the resonator. The resonator
was optimized for fast readout. By studying quasiparticle tunneling
events in the SCB, we have characterized the performance of the
readout and found that we can perform a single shot parity measurement
in approximately 50 ns. This is an order of magnitude faster than
previously reported measurements.
\end{abstract}


\maketitle

\section{Introduction}

Superconducting devices based on Josephson junctions have successfully
been used in many different kinds of applications, including very sensitive
magnetometers based on superconducting interference devices (SQUIDs),
bolometric detectors, mixers, and parametric amplifiers. 
They have also been suggested to be strong candidates as building
blocks for a quantum computer
\cite{Makhlin:RMP,Platenberg:FluxCNOT,McDermott:SimMeas,Koch:Transmon}.
They are easily fabricated with standard lithographic techniques and
can be integrated with other electrical circuits. This gives them the
potential to be scalable. Devices utilizing charging effects include
very small tunnel junctions. Such Coulomb blockade devices, are also
widely used in measurements, for example, the single electron
transistors (SET)\cite{Fulton:SET}. The radio frequency version of
the SET is the worlds most sensitive electrometer
\cite{RF-SET,Brenning:RF-SET}.

The single Cooper-pair box (SCB)
\cite{Buttiker:SCB,Bouchiat:SCB,Nakamura:SCB}
is one of the simplest Coulomb blockade devices, involving a single
Josephson junction. SCBs are very sensitive to the presence of
quasiparticles which suggested its use as potential radiation detector
\cite{Shaw:Detector}.
The presence of quasiparticles can be measured by detecting the
charge on the SCB island using an external SET
\cite{Ferguson:QuasiParticles}.
In this paper, we characterize an intrinsic method for reading out
the SCB which relies on the curvature of its energy bands. This
methods is both faster and is predicted to have less
backaction then using a SET. The curvature of the energy bands of the
SCB (with respect to gate charge) gives rise to the so called quantum
capacitance \cite{Duty:QC,Silanpaa:JC} and has been utilized in a
number of experiments, for example in the measurements on longitudinal
dresses states of a driven SCB
\cite{Wilson:LDS,Wilson:DressedRelaxation}. It has also been
used to study the ground state of two coupled qubits
\cite{Shaw:Entangled}, and to study quasiparticle poisoning of a SCB
qubit \cite{Shaw:QuasiParticles}. In
Ref.~\onlinecite{Shaw:QuasiParticles} they used a resonator with a
bandwidth of 200 kHz which limited the speed the their measurements.
Here we use the random tunneling of quasiparticles to characterize the
performance of the quantum capacitance readout. We show that we can
measure the state of the SCB an order of magnitude faster than has
previously been reported, including measurements using RF-SETs. We
also show that we can prepare the SCB in a certain parity state with
a high probability.

This paper is structured as follows: in section~\ref{sec:theory}, we
present the theory behind the readout technique of using the quantum
capacitance. In section~\ref{sec:fabrication}, we show how the samples
were designed and fabricated and then in
section~\ref{sec:measurements} we present the measurements done to
characterize the readout.
 
\section{Theory}\label{sec:theory}

\begin{figure}[htb!]
  \includegraphics[width=\columnwidth]{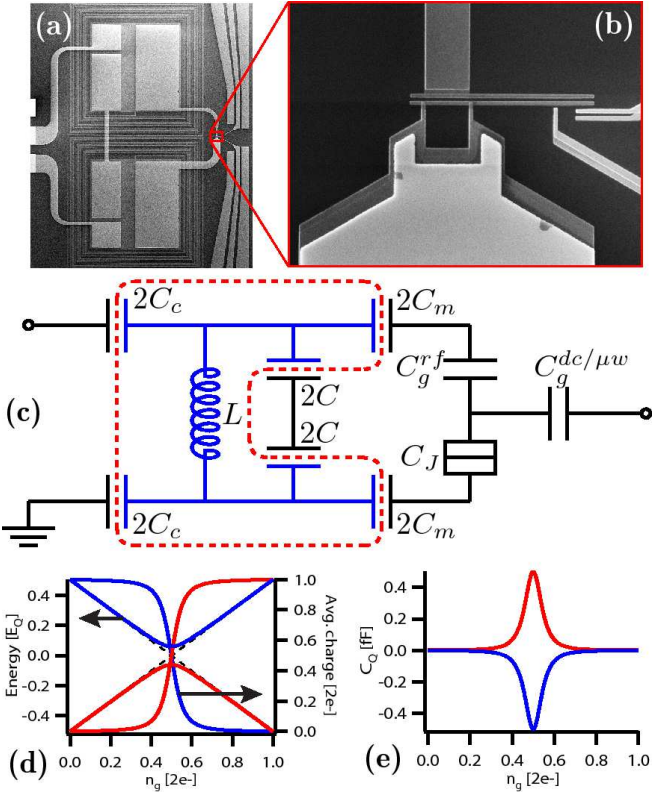}%
  \caption{\label{fig:device} (Color online)
    (a) A scanning electron micrograph of the resonator for sample
    A. The inductor can be seen spiraling around the plates of the
    capacitors with the upper part counter wound compared to the
    bottom.
    (b) A scanning electron micrograph of the SCB for sample A. The
    SCB consists of a 5 $\mu$m long and 100 nm wide superconducting
    island. The SCB for sample B is the same except that the island is
    8 $\mu$m long in order to increase the gate capacitance. The
    island is connected to a reservoir through two small Josephson
    junctions in a SQUID geometry. The potential of the island is
    controlled by three different capacitive gates. One large rf gate,
    $C_g^{rf}$, (on the top) is connected to the resonator and used
    for the readout. In addition, there is a dc gate, $C_g^{dc}$, used
    to bias the SCB at the working point and a microwave gate,
    $C_g^{\mu w}$, used for spectroscopy.
    (c) A schematic of the device. The parallel resonator
    consists of two metal layers separated by a thin insulating
    layer. The bottom layer is of Nb (within the dashed line),
    forming the inductor and bottom plates of the capacitors. Next is
    a 200 nm layer of silicon nitride (the dashed line) covering the whole
    sample and, finally, the top layer is Al (outside the dashed line)
    making the top plates of the capacitors as well as the SCB.
    (d) The energy of the two lowest energy eigenstates of the SCB, as
    well as the expectation value of the (excess) charge on the SCB island
    for each state as a function of the normalized gate voltage,
    $n_g$.
    (e) The quantum capacitance, $C_Q$, for the two eigenstates as a
    function of the normalized gate voltage, $n_g$.
  }
\end{figure}

\subsection{Cooper-pair box}
The sample under investigation is a SCB and is shown in
Fig.~\ref{fig:device}(a) \& (b). The SCB consist of a superconducting
island connected to a large reservoir by a Josephson junction. The
Josephson junction is made in a SQUID-configuration to allow the
Josephson energy, $E_J$, to be tuned by applying a magnetic field
through the loop. The SCB is also characterized by the electrostatic
energy, $E_{el} = E_Q(n-n_g)^2$, which is the energy required to add
an extra Cooper pair to the island. Here $n$ is the number of Cooper
pairs that have tunneled onto the island and $n_g = C_gV_g/2e$ is the
normalized gate voltage, where $C_g$ is the capacitance between the
voltage source and the island.
If the capacitance of the island, $C_\Sigma$, is small enough the
Cooper-pair charging energy, $E_Q = (2e)^2/2C_\Sigma$, will dominate
over the Josephson energy, $E_J$, and the temperature, $k_B T$.
In this case the charge fluctuation on the island will be small. The
number of excess Cooper pairs on the island, $n$, is then a good
quantum number and the charge of the island can be well controlled by
the external gate voltage, $V_g$. For $E_J \ll E_Q$ and $0 < n_g < 1$,
only two charge states will be of interest: $\left|0\right>$ and
$\left|1\right>$, corresponding to zero ($n=0$) or one ($n=1$) extra
Cooper pair on the island. Then the Hamiltonian of the Cooper-pair box
can be written
\begin{gather}\label{eq:Hcpb}
  H = -\frac{1}{2}E_Q(1-2 n_g) \sigma_z -\frac{1}{2}E_J \sigma_x
\end{gather}
where $\sigma_x$, $\sigma_z$ are the Pauli spin matrices. Here we have 
ignored all state independent terms of the Hamiltonian. The two
eigenenergies for this Hamiltonian are plotted in
Fig.~\ref{fig:device}(d) as a function of $n_g$. At the degeneracy
point $n_g = 0.5$, where the electrostatic energies of the two charge
states cross, we get an avoided level crossing with a splitting
between the ground and excited state equal to $E_J$. In
the same graph, we have also plotted the expectation value of the
island charge $\left<n\right>$ for each energy eigenstate.

\subsection{Quantum Capacitance}
At the degeneracy point the energy difference between the ground and
excited state is, to first order, independent of the gate charge $n_g$,
which makes this the ideal bias point when the SCB is used as a qubit.
At this point, the longest  dephasing times are obtained
\cite{Vion:Quantronium}. Therefore, this is often called the optimal
point. If we want to detect the state of the SCB sitting at the
optimal point, we cannot measure the charge, since the charges of the
ground and exited state are the same at this point (see
Fig.~\ref{fig:device}(d)). Although the charges are the same for the two
states, the derivatives of the charges with respect to the gate voltage
differ and can be used for readout. We can define an effective
capacitance of the SCB by calculating $C_{eff} = \partial
\left<Q_g\right>/\partial V_g$, where $\left<Q_g\right> =
C_gC_jV_g/C_{\Sigma} + 2e\left<n\right>C_g/C_{\Sigma}$ is the average
value of the injected charge on the gate capacitor. This effective
capacitance, $C_{eff} = C_{geom} + C_Q$, will have two contributions,
a geometric part $C_{geom} = C_g C_J/(C_g+C_j)$ consisting of the gate
capacitance in series with the junction capacitance, and a state
dependent part that we call the quantum capacitance
\cite{Duty:QC}. This quantum capacitance, $C_Q$, takes the following
form
\begin{gather}
  C_Q^{g/e} = \pm \frac{C_g^2}{C_\Sigma}
  \frac{\alpha^2}{\left(\alpha^2+(1-2 n_g)^2\right)^{\nicefrac{3}{2}}}
  = \pm \frac{C_g^2}{C_\Sigma} \frac{E_Q E_J^2}{\Delta E^3}
\end{gather}
for the ground (g) and first excited (e) state, where $\alpha =
E_J/E_Q$ and $\Delta E = \sqrt{E_J^2+E_Q^2(1-2n_g)^2}$
is the energy difference between the two states at a given $n_g$. $C_Q$
is equal in magnitude but has opposite signs for the ground and
excited state, with $C_Q$ being negative for the ground state. In
Fig.~\ref{fig:device}(e), $C_Q$ is plotted against the gate charge,
$n_g$, for the ground and exited state of the SCB. 

If we embed the SCB in a resonant tank circuit (see
Fig.~\ref{fig:device}(c)) the effect of the quantum capacitance will be
to shift the resonance frequency. The reflection coefficient, $\Gamma =
|\Gamma|\exp(\phi)$, of the resonator has a constant magnitude,
{\it i.e.} $|\Gamma| = 1$, since there are no dissipative element in
the resonator. The phase $\phi$, however has a sharp frequency
dependence and close to the resonance frequency of the tank circuit,
$\omega_0 = 2\pi f_0 = 1/\sqrt{L(C+C_c)}$, the phase, $\phi$, can be
approximated by the expression:
\begin{gather}
  \phi = -\pi - 2\arctan \left( 2Q\frac{\omega-\omega_0}{\omega_0}
  \right),
\end{gather}
where $\omega = 2\pi f$ is the angular probe frequency and $Q =
(C+C_c)/C_c^2Z_0\omega_0$ is the external quality factor of the
resonator. Now, if the capacitance of the SCB is changed by a magnitude
of $\Delta C_Q$, the phase of the rf signal at frequency $f_0$
reflected off the resonator will change by:
\begin{gather} \label{eq:phase_shift}
  \Delta\phi \approx -2\arctan\left( Q\frac{\Delta C_Q}{C+C_c} \right).
\end{gather}
Here we have treated the resonator purely classically. A full
quantum treatment, including the SCB, resonator and transmission line
can be found in the work of Johansson
{\it et al.} \cite{Johansson:QCReadout}. It is shown that this method
of readout is quantum limited \cite{Johansson:QCReadout}, meaning that
no information is lost during the readout (no extra dephasing).

\subsection{Quasiparticles}


Quasiparticles are single-particle excitations of the superconducting
condensate. Quasiparticle fluctuations have been studied as a source
of noise in superconducting devices for some time.  Thermodynamic
fluctuations in the quasiparticle number, also know as
generation-recombination noise, is an important source of noise at
intermediate temperatures \cite{Wilson:QP-noise}. Time-resolved
measurements of these fluctuations have shown very good agreement
between theory and experiment
\cite{Wilson:QP-numfluc,Wilson:QP-meas}. At very low temperatures,
were thermal quasiparticles should be suppressed, a significant
population of quasiparticles is still observed in most
experiments. The origin of these nonequilibrium quasiparticles is
still unknown. However, it is clear that they remain an important
source of noise.

The most significant source of quasiparticle noise in a SCB is
commonly referred to as quasiparticle poisoning.  When a
nonequilibrium quasiparticle in the reservoir tunnels onto the SCB
island, it shifts the potential of the island by $\pm e/C_{\Sigma}$.
The random tunneling of quasiparticles on and off of the island,
therefore leads to a large-amplitude telegraph noise
in the island potential. Quasiparticle poisoning has been extensively
studied
\cite{Shaw:QuasiParticles,Aumentado:QuasiParticles,Lutchyn:QuasiParticles},
including the observation of individual tunneling events in real time
\cite{Ferguson:QuasiParticles,Naaman:QuasiParticles}.
It has been shown that poisoning is well described by a simple kinetic
model, starting from the assumption of a finite density of
nonequilibrium quasiparticles in the leads.  In this model, the
tunneling rates are then dictated by the relative energies of the even
(no quasiparticles on the island) and odd (one quasiparticle on the
island) states.  If the energy of the odd state is lower than the even
state, the quasiparticle can be trapped on the island (see
Fig.~\ref{fig:measurements}(a)).  The energy difference of the even
and odd state is maximum at the charge degeneracy point, $n_g = 0.5$,
where it takes the value $\delta E =
E_Q/4-E_J/2-\Delta_i+\Delta_r$. Here $\Delta_i$ and $\Delta_r$ are the
superconducting energy gap of the island and the reservoir,
respectively. The average occupation of the two states is then simply
determined by the Boltzmann factor $\exp(-\delta E/kT)$.  

\section{Device design and fabrication}\label{sec:fabrication}

When designing the readout circuit there is a trade off between
having a good signal (large phase shift) and a low backaction. The
phase shift is roughly proportional to $QC_Q/(C+C_c)$ (see equation
\ref{eq:phase_shift}), meaning that for a fixed Q-value ({\it i.e.}
measurement bandwidth) we want a small total capacitance in the
resonator (and a large inductance). However, the voltage noise of the
environment will induce charge fluctuations on the SCB, with a
magnitude that scales with the same prefactor.
Thus, by decreasing the total capacitance in the resonator, $C+C_c$,
you can increase the sensitivity (larger phase shift) at the cost of
a larger backaction on the SCB.

The Q-value of the resonator sets the bandwidth, {\it i.e.} an upper
limit on how fast we can measure. We designed the resonator to have an
external Q-value of 100 which corresponds to a time constant of 50 ns
at $f_0 = 650$ MHz. The internal Q-value is usually substantially
larger and can therefore be ignored.

The devices were fabricated in a multilayer process. Starting from a
high-resistivity silicon wafer with a native oxide, the wafer was
first cleaned using rf back sputtering directly after which a 60 nm
thick layer of niobium was sputtered. To pattern the niobium, we used
a 20 nm thick Al mask made by e-beam lithography and e-beam
evaporation. The niobium was then etched in a CF$_4$ plasma (with a
small flow of oxygen) to form the inductor and bottom plates of the
capacitors (see Fig.~\ref{fig:device}(c)). The choice of niobium for
the bottom layer made it possible to test the resonator in liquid
helium (even with a top layer made of normal metal, {\it e.g.}
gold). The Al mask was removed with a wet-etch solution based on
phosphoric acid. Before depositing the insulator we cleaned the wafer
in a 2\% HF solution for 30 sec in order to remove most of the niobium
oxide which has been found to degrade the Q-value of the niobium
resonators. Using PE-CVD, we then deposited an insulating layer of
200\,nm of silicon nitride. The silicon nitride layer covers the whole
wafer; connections to the niobium layer are only made through
capacitors. We chose silicon nitride since it is known to have low
dielectric losses \cite{Martinis:DielectricLoss}. After using a
combination of DUV photolithography to define bonding pads along with
e-beam lithography to define quasiparticle traps
\cite{Rooks:DUV-EBL}, a 3/80/10 nm thick layer of Ti/Au/Pd was
deposited by e-beam evaporation. Finally, the layer containing the SCB
was made by e-beam lithography and two-angle shadow evaporation of
10+30 nm of aluminum, with 6 min of oxidation at 4 mbar. The thickness
of the island (10 nm) was chosen to be much thinner than for the
reservoir (30 nm) in order to enhance the superconducting gap,
$\Delta_i > \Delta_r$, of the island compared to the reservoir. This
was done to reduce quasiparticle poisoning.

\section{Measurements}\label{sec:measurements}

\subsection{Measurement setup}

\begin{figure}[htb!]
  \includegraphics[width=0.7\columnwidth]{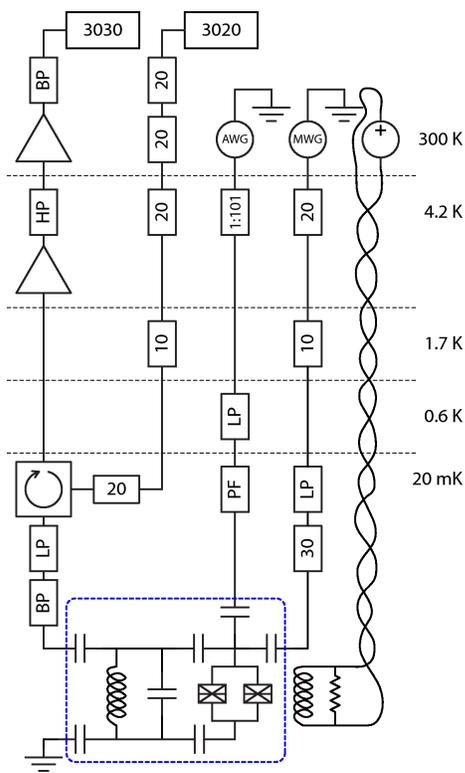}%
  \caption{\label{fig:setup} (Color online)
    The measurement setup used in the experiment. The sample
    (inside the dashed line) is mounted inside a copper box on the
    mixing chamber of a dilution refrigerator, which reaches a
    temperature of 20 mK. The readout pulse is created using a
    Aeroflex 3020 signal generator (with built in IQ modulation). It
    is heavily attenuated and coupled through a circulator located at
    the mixing chamber and filtered through two RLC waveguide filters
    before reaching the resonator. On the way up, the signal is first
    amplified at 4.2K with a Quinstar amplifier with noise temperature
    of 1 K and then amplified again at room temperature with a
    Mini-Circuit amplifier. Before the signal is finally digitized
    using an Aeroflex 3030 vector analyzer (with a 33 MHz bandwidth),
    it is first filtered using an RLC bandpass filter to reject image
    noise.
  }
\end{figure}

The device was cooled in a dilution refrigerator with a
base temperature of about 20 mK. For the readout, we used an Aeroflex
3020 signal generator to produce the rf signal. The signal was heavily
attenuated and filtered and was fed to the tank circuit via a
Pamtech circulator positioned at the mixing chamber. The reflected
signal was amplified by a Quinstar amplifier at 4K with a nominal
noise temperature of 1 K. The in-phase and quadrature components of
the signal were finally measured using an Aeroflex 3030 vector
digitizer. A schematic of the measurement setup can be found in
Fig.~\ref{fig:setup}.

\subsection{Device characterization}

In the following sections we will show measurements of two different
devices, referred to as sample A and sample B. In order to
characterize the devices and extract parameters, the resonator was
first measured using a network analyzer. From the measured phase
response we could extract the resonance frequency, $f_0$, and the
Q-value of the resonator. We could extract both the Cooper-pair
charging energy, $E_Q$, and the maximum Josephson energy,
$E_J^{max}$ by conventional spectroscopy, while applying a
perpendicular magnetic field through the SQUID loop of the
SCB, and thereby tuning $E_J$. Finally by measuring $C_Q$ as a
function of gate charge, we can extract the rf gate capacitance and
the total resonator capacitance. From the expression for the Q-value
and the resonance frequency, we can then extract the parameters for the
individual components of the resonator. The extracted parameters for
the two devices are presented in Table~\ref{tab:params}. The
values of the parameters for sample B corresponds reasonable well to the
geometrically identical device that was used in
Ref.~\onlinecite{Persson:Sisyphus}. The gate capacitance for sample A
roughly agrees with what you would expect from the 5 $\mu$m island
compared the to the capacitance of the 8 $\mu$m island in sample B. If
we the insert the parameters from Table~\ref{tab:params} into 
Eq.~(\ref{eq:phase_shift}) we calculate an expected phase shift of 11
deg for sample A and 30 deg for sample B between the two parity states
when biased at the degeneracy point.

\begin{table}
\caption{\label{tab:params} The extracted parameters for the two
  samples.
}
\begin{ruledtabular}
\begin{tabular}{ccc}
& Sample A & Sample B \\
\hline
$f_0$ & 676 MHz & 663 MHz \\
$Q$ & 128 & 130 \\
$L$ & 151 nH & 324 nH \\
$C$ & 251 fF & 97 fF \\
$C_C$ & 116 fF & 81 fF \\
$C_g^{rf}$ & 0.2 fF & 0.3 fF \\
$E_Q$ & 62 GHz & 48 GHz \\
$E_J^{max}$ & 7.2 GHz &  7.4 GHz \\
\end{tabular}
\end{ruledtabular}
\end{table}

\subsection{Readout performance}

\begin{figure}[htb!]
  \includegraphics[width=0.7\columnwidth]{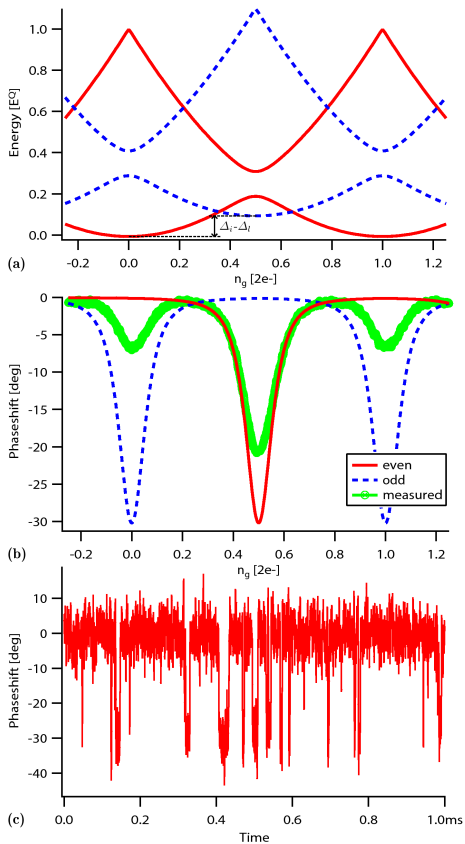}%
  \caption{\label{fig:measurements} (Color online)
    (a) The energy as a function of $n_g$ for the even (solid) and odd
    (dashed) parity  of the SCB island. The odd state energy is
    shifted by the difference in the superconducting gaps of the
    island and the leads, $\Delta_i-\Delta_l$. If  $E_Q/4-E_J/2 >
    \Delta_i-\Delta_l$ the island will form a trap for quasiparticles
    at $n_g = 0.5$. 
    (b) The phase shift as a function of the gate charge, $n_g$, for
    the even (solid) and odd (dashed) state. We also show the average
    measured phase for sample B, which shows contributions from both
    parity states, while repetitively sweeping the dc gate at a
    repetition rate of 5 kHz. The rate of the dc sweep was fast
    compared to the quasiparticle tunneling rate, such that the
    quasiparticle tunneling probability during one period was low. The
    SCB then spent most of the time in the even state, which has on
    average a lower energy than the odd state.
    (c) A typical time trace of the measured phase, from an identical
    device to sample B, when sitting at the even degeneracy point,
    $n_g = 0.5$. The SCB spends most of the time in the odd state
    ($\phi = 0$ deg) but makes short excursions to the even state
    ($\phi = -30$ deg).
  }
\end{figure}

We have fabricated and measured a number of samples. All devices we
have measured so far have been poisoned, meaning that
quasiparticles cause switching between the two parity states of having
an even or odd number of quasiparticles on the island. The switching
between these parities happens on the time scale of a few
microseconds. Although this is far from ideal for many applications,
it has given us a way to characterize the readout. When a
quasiparticle tunnels on or off the island the quantum capacitance of
the SCB will change, and can thus be detected as a change in the
reflected phase from the resonator. The measured time averaged phase
for sample B is shown in Fig.~\ref{fig:measurements}(b) as function of
$n_g$. In Fig.~\ref{fig:measurements}(c) we show the time dependence
of the phase measured at the degeneracy point, $n_g = 0.5$. Most of the
time the SCB is in the odd state with a phase shift of about 0
degrees, however now and then the extra quasiparticle escapes the
island and the SCB spends short periods in the even state with a phase
shift close to -30 degrees. 

\begin{figure}[htb!]
  \includegraphics[width=\columnwidth]{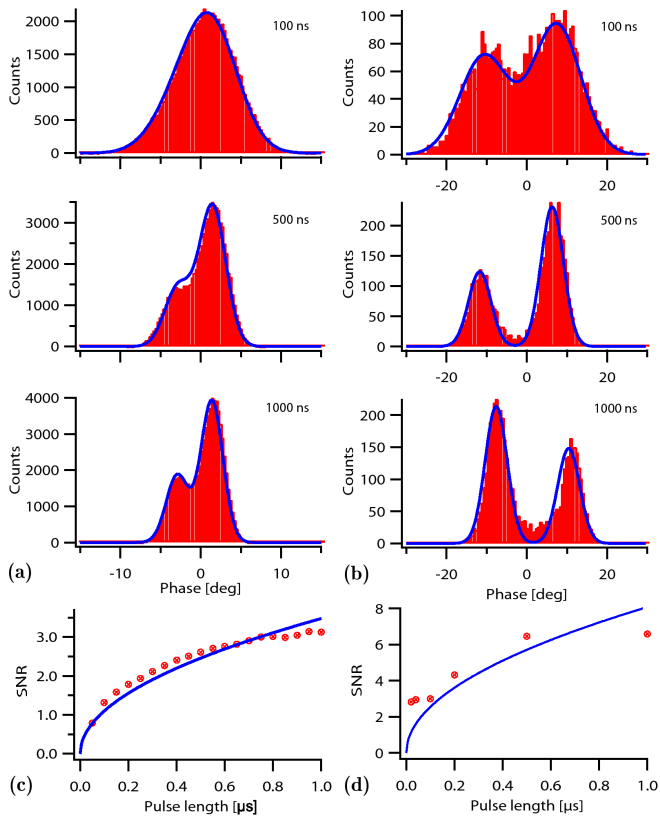}%
  \caption{\label{fig:data} (Color online)
  Sitting at $n_g = 0.5$ we perform pulsed measurements and extract
  the phase of each pulse. For three different pulse lengths we repeat
  the measurements many times and make histograms of the phases
  (bars). In (a), we show histograms for measurements on sample A and
  in (b) for sample B. Sample A was designed to have a larger total
  capacitance for the resonator and a smaller rf-gate capacitance than
  in sample B. This reduces the backaction of the measurement at the
  expense of lowering the SNR. To the histograms, we fit a double
  Gaussian (solid line) with the same standard deviation for the two
  peaks. We then define the SNR of the measurement as the peak
  separation divided by the standard deviation. The above procedure is
  repeated for different pulse lengths, ranging from 20 ns to 1
  $\mu$s. The SNR is then plotted as function of the pulse length for
  sample A in (c) and for sample B in (d). We see that the SNR roughly
  follows the expected square root dependence on the measurement
  time. Since sample A was designed to have a lower backaction, we
  expect it to have a lower signal. The calculated phase shift is 11
  deg for sample A and 30 deg for sample B. This value is however
  calculated for very low probe powers. In these measurements, the
  phase shift is reduced by the relative large measurement power.
  }
\end{figure}

While sitting at the degeneracy point, $n_g = 0.5$, we have also
performed pulsed measurements of the state. We send down a Gaussian
pulse, with a length defined as the full width at
half maximum (FWHM), and measure the phase of the reflected pulse. In
order to optimize the signal to noise of the measured response, we used
a so-called matched filter, where the time traces of the measured
in-phase and quadrature component (I and Q) are multiplied with a
Gaussian template (with the same shape as the one generated by the
signal generator). The product is then integrated and a single value
for is extracted. This is done for both I and Q and we can calculate
the phase by $\phi = \arctan\left(Q/I\right)$.
We perform 100 000 of these measurements and make histograms of the
measured phase and see two peaks centered at different phases
corresponding to the two parity states. We fit a double Gaussian (both
with the same standard deviation) to the histograms (see
Fig.~\ref{fig:data}(a,b)). We define the signal-to-noise ratio
(SNR) of the measurement as the peak separation divided by the
standard deviation. This is performed for different pulse lengths and
we extract the SNR as a function of the measurement time,
Fig.~\ref{fig:data}(c,d). The SNR roughly follows the expected square
root dependence on the measurement time. To reach a SNR $\gtrsim 1$ we
need a pulse length of the order of 50-100 ns in both samples. The
shortest measurement time was here limited by the time constant of
resonator which was about 60 ns for both of the samples. 

\subsection{Quasiparticle relaxation and state preparation}

\begin{figure}[htb!]
  \includegraphics[width=0.7\columnwidth]{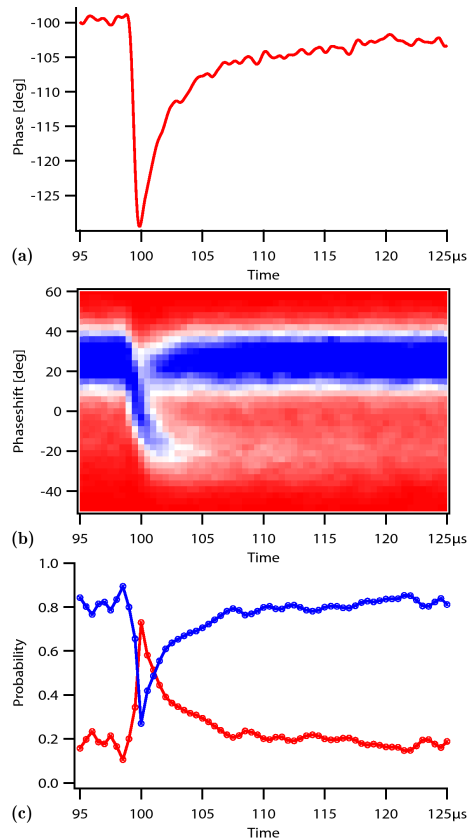}%
  \caption{\label{fig:qphistograms} (Color online)
    Measurement of the quasiparticle relaxation in sample B.
    (a) The average of 1000 traces of the measured phase 
    after pulsing the gate from the odd degeneracy point ($n_g = 0$)
    (at time 100 $\mu$s) to the even degeneracy point ($n_g = 0.5$)
    and continuously monitoring the phase as a  function of time. 
    (b) Histograms of the measured state as a function of time. The
    measured phase for each repetition is divided into 0.5 $\mu$s
    increments. We then make a histogram of the average phase for each
    increment, measured from the start of the pulse. This produces a
    histogram as a function of time.
    (c) Occupation probability of the even (lower) and odd (upper)
    state as function of time extracted from the histograms
    in b. After pulsing to the even degeneracy point, the
    probabilities are inverted suggesting that we should be able to
    prepare the system in the even state with high probability even if
    the device has significant quasiparticle poisoning.
  }
\end{figure}

We know that by sitting at the even degeneracy point the system will
eventually relax into the odd parity state given that 
$E_Q/4-E_J/2 > \Delta_i-\Delta_l$ (see Fig.~\ref{fig:measurements}). 
We wanted to study how fast this process is and to what extent you
can prepare the SCB in a certain parity state. We start by letting
the system equilibrate biased at the odd degeneracy point, {i.e. $n_g
  = 0$}, thereby preparing the even state. We then pulse the gate to
the even degeneracy point (at $t=100$ $\mu$s) and observe the
dynamics.
From a long time trace, including 1000 pulses, we divide each
repetition into 0.5 $\mu$s increments. We extract the average phase
from each increment and each repetition. We then make a histogram of
the phase for each increment as a function of time  
(see Fig.~\ref{fig:qphistograms}(b)). We fit the histograms to a double
Gaussian and extract the occupation probability of the even and odd
state as a function of time (see
Fig.~\ref{fig:qphistograms}(c)). Sitting a the odd degeneracy
point, $n_g = 0$, there is an equilibrium probability of more than
$80\%$ of being in the even state and when we pulse to the even state
most of the probability is preserved. Eventually the system
equilibrated and then the probability is reduced to $~20\%$. From the
relaxation of the probabilities we extract an equilibration time of
~2.8 $\mu$s. As a comparison, we show the average time trace (see
Fig.~\ref{fig:qphistograms}(a)) where we have
taken the average of the full time traces from different pulses. From
this we extract a relaxation rate of ~3 $\mu$s in good
agreement. Since the quasiparticle relaxation rate is much longer
than the operation and readout times this suggests that, even if the
device is poisoned, we should with high probability be able to prepare
the box in the even state and perform useful measurements.

\section{Conclusions}

We have fabricated and tested two samples with a SCB together with an
on-chip lumped element resonator. The resonators were optimized for
readout speed, with a Q-value around 100. We have characterized the
readout by employing the effect of quasiparticle poisoning and found
that for readout pulses of length 50-100 ns we get a SNR greater then
1. 

\begin{acknowledgments}
We thank the members of the Quantum Device Physics and Applied Quantum
Physics groups for useful discussions. The samples were made at the
nanofabrication laboratory at Chalmers. The work was supported
by the Swedish VR and SSF, the Wallenberg foundation, the EU
under the project EuroSQIP and by IARPA through ARO award
W911NF-09-1-0376.
\end{acknowledgments}


\end{document}